**Parameter extraction of Extended Floating Gate Field Effect Transistors (EGFETs): Estimating the threshold voltage, series resistance, and mobility degradation from I-V measurements**

Authors:

Yunsoo Park and Santosh Pandey[*]

Electrical and Computer Engineering, Iowa State University, Ames, Iowa, USA 50011.

Corresponding Author: [*]Santosh Pandey

Email: pandey@iastate.edu ; yunsoopk@iastate.edu

Abstract:

Extended Floating Gate Field Effect Transistors (EGFETs) are CMOS-compatible floating gate devices capable of detecting charges on their sensing area by the relative shifts in current-voltage (I-V) characteristics. The I-V shifts are generally computed by measuring the EGFET parameters in the strong inversion region of operation. This could lead to errors in estimating the device sensitivity because the simple I-V model ignores the mobility degradation and series resistance effects in EGFETs. Our goal is to model these parasitic effects and present methods to extract the key device parameters. We derive an analytical I-V model for EGFETs in the linear region of transistor operation, accounting for both the mobility degradation and series resistance effects. Based on the analytical model, three methods are presented to estimate the key parameters, namely the threshold voltage, series resistance, surface roughness parameter, low-field mobility, and effective mobility from the I-V characteristics, gate transconductance, and drain conductance. The peak transconductance method is used as a benchmark for comparing the extracted threshold voltages. Silicon-based EGFET devices are fabricated, and their I-V characteristics are measured with deionized water and three polyelectrolytes. From the I-V data, the parameter extraction methods are used to compute the values of the key parameters, and the suitability of each method is discussed. The gate transconductance methods show good agreement between the values for the key parameters, while the drain transconductance method gives lower values of the key parameters. There is scope to improve the presented methods by incorporating the effects of substrate bias and asymmetric series resistance for new extended-gate device architectures, including solution-based organic field-effect transistors.

## 1. Introduction

Extended Floating Gate Field Effect Transistors (EGFETs) have been used to detect the chemical and biological properties of specimens, including gases, charged electrolytes, polymers, protein, and DNA strands [1–4]. The main benefit of EGFETs lies in its CMOS-compatibility and on-chip integrability with signal processing circuitry. The EGFET floating gate is capacitively coupled to a sensing area on one side and to a control gate on the other side. Charged molecules placed on the sensing area produce a shift in the EGFET current-voltage (I-V) characteristics, which is represented by a current or voltage sensitivity index. The current sensitivity index is computed as the difference between two drain currents $I_{ds}$ at a fixed drain voltage in the saturation region with respect to the ionic charge (e.g., $\Delta I_{ds} / \Delta pH$). For the voltage sensitivity index, the threshold voltage $V_T$ of EGFETs is extracted by extrapolating the square root of drain current $I_{ds}$ versus gate voltage $V_{gs}$ curve to zero gate voltage (called the linear extrapolation method), and thereafter the shift in threshold voltage $V_T$ is computed with respect to the ionic charge (e.g., $\Delta V_T / \Delta pH$) [5].

While the above approaches to compute the EGFET current or voltage sensitivity have been adopted for their simplicity, there are reasons to investigate alternative methods for the accurate parameter extraction of EGFETs, including its threshold voltage $V_T$. Firstly, in the strong inversion region, EGFETs (like other transistors) are prone to mobility degradation which may not produce linear I-V curves for the linear extrapolation method. As such, the linear or subthreshold regions of EGFET operation are more suitable for parameter extraction. Secondly, besides the linear extrapolation technique, other $V_T$ extraction methods based on the gate transconductance $g_m$ or a combination of $I_{ds}$ and $g_m$ have been suggested in the MOSFET community which closely reflect the tipping point where the surface potential is twice the Fermi potential [6,7]. A comparative analysis of other $V_T$ extraction methods and their suitability has largely been unexplored for EGFET devices. Thirdly, EGFETs have series resistance at the source and drain terminals ($R_{SD}$) whose value may be significant in devices with long lead lines or poor source/drain contacts. The effects of series resistance $R_{SD}$ on the I-V characteristics are often ignored in EGFET measurements [8,9], but need to be incorporated to gauge the parasitic effect of $R_{SD}$ on EGFET performance.

In this work, our goal is to extract the key parameters of EGFETs from I-V characteristics of a single device. We focus on extraction methods of the following key parameters – threshold voltage $V_T$, low-field mobility $\mu_0$, mobility degradation (or surface roughness) parameter $\theta$, effective mobility $\mu_{eff}$, and series resistance $R_{SD}$. However, it is difficult to extract the low-field mobility $\mu_0$ from the drain current $I_{ds}$ or gate transconductance $g_m$ alone using closed-form analytical models. This is because the mobility degradation effect is embedded in both the $I_{ds}$ and $g_m$ expressions. To circumvent this issue, our approach is to isolate the surface roughness parameter $\theta$ in the analytical models before extracting the threshold voltage $V_T$ and low-field mobility $\mu_0$. In Section 2, we describe the layout and schematics of the fabricated EGFETs, including the electrical characterization methods. In Section 3.1, we derive analytical expressions of the drain current $I_{ds}$, gate transconductance $g_m$ (= $\partial I_{ds} / \partial V_{gs}$) and drain transconductance $g_{ds}$ (= $\partial I_{ds} / \partial V_{ds}$) incorporating the surface roughness parameter $\theta$ and source-drain series resistance $R_{SD}$. The series resistance $R_{SD}$ is estimated from the measured output resistance of the EGFETs. Thereafter, the peak transconductance method [6] (Section 3.2) is used to obtain the threshold voltage which serves as a benchmark for the following $V_T$ extraction methods: (i) In the first method [10] (Section 3.3), the ratio of drain current $I_{ds}$ to square root of gate transconductance $g_m$ (i.e., $I_{ds} / g_m^{0.5}$) is plotted with respect to the gate voltage $V_{gs}$ to obtain the threshold voltage $V_T$ and low-field mobility $\mu_0$ from the x-intercept and slope, respectively. Then the surface roughness parameter $\theta$ and effective mobility $\mu_{eff}$ are extracted from the drain current $I_{ds}$ and gate transconductance $g_m$. (ii) In the second method (Section 3.4), the second derivative of the reciprocal of the drain current (i.e., $\partial^2 (1/I_{ds}) / \partial V_{gs}^2$) is plotted with respect to the gate voltage $V_{gs}$ to give the threshold voltage $V_T$ and low-field mobility $\mu_0$ from the x-intercept and slope, respectively. The surface roughness parameter $\theta$ and effective mobility $\mu_{eff}$ are then extracted. (iii) In the third method (Section 3.5), the ratio of drain transconductance $g_{ds}$ to its derivative (i.e., $g_{ds} / (\partial g_{ds}/ \partial V_{gs})^{0.5}$) with respect to the gate voltage $V_{gs}$ is plotted to give the threshold voltage $V_T$ and low-field mobility $\mu_0$ from the x-intercept and slope, respectively. The surface roughness parameter $\theta$ and effective mobility $\mu_{eff}$ are extracted from the known values of $V_T$ and low-field mobility $\mu_0$. In Section 3.6, the parameter extraction methods are tested with three charged polyelectrolytes on the sensing area of the EGFETs. In Section 4, we discuss the implications of our results and the suitability of presented methods for the parameter extraction of EGFETs. It is noted that some of these methods presented here have been developed for MOSFET devices but have not been tested on EGFET devices for their suitability.

## 2. Materials and Methods

Fig. 1a shows the circuit representation and schematic of EGFETs fabricated on p-type silicon substrate using the 1.5 μm dual-poly CMOS fabrication process. The width to length ratio is 4.5 μm / 1.5 μm. The gate oxide thickness is 30.9 nm and the inter-poly oxide thickness is 60 nm. The intrinsic gate-to-source and drain-to-source voltages are denoted as $V_{GS}$ and $V_{DS}$, respectively. The applied gate-to-source and drain-to-source voltages are denoted as $V_{gs}$ and $V_{ds}$, respectively. The source and bulk contacts are grounded. The source and drain resistances are $R_s$ and $R_d$, respectively, while the total source-to drain is $R_{sd}$. We assume that $R_d = R_s = R_{sd} / 2$. The series resistances arise from the source/drain contacts, electrical wires, and pad contacts. The total charge on the gate sensing electrode is labelled as $Q_{gs}$. Fig. 1b shows an image of the fabricated EGFET devices packaged as a 40-pin, 2.2 mm × 2.2 mm chip.

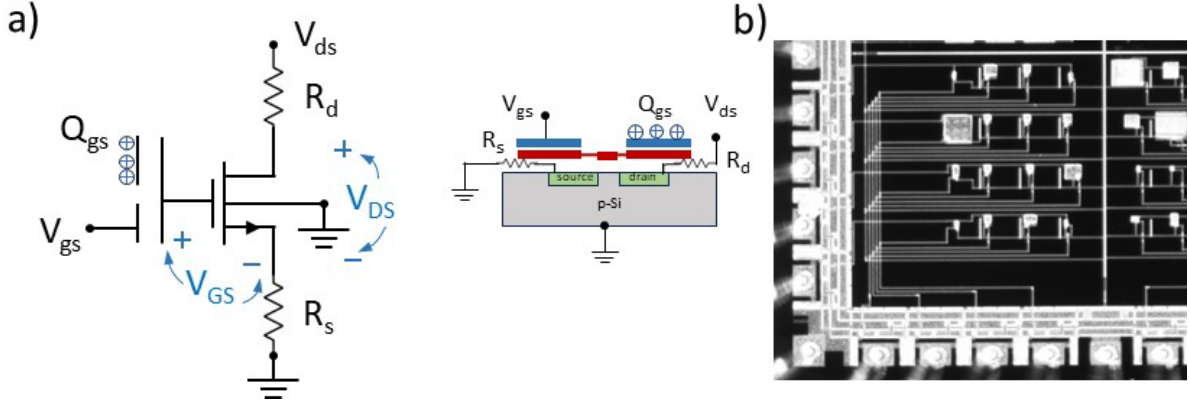

**Fig. 1.** Device schematic and fabricated EGFET chip. (a) EGFETs are tested in common-source configuration in the linear region of transistor operation. The external parasitic resistances at the source and drain are $R_s$ and $R_d$, respectively. The external voltage at the gate-source and drain-source terminals are $V_{gs}$ and $V_{ds}$, respectively. (b) Sectional image of the fabricated 40-pin chip showing the EGFET devices, metal lines, and the contact pads.

Electrical measurements of EGFETs were conducted using the B1500A Semiconductor Device Parameter Analyzer (Keysight, Colorado Springs, CO, USA). LabView programs were written for the semiconductor parameter analyzer to automate the I-V data acquisition in 0 to 4V voltage range with 100 mV step size. The I-V measurements were conducted with different solutions on the sensing area: deionized (DI) water, 0.5 % poly aspartic acid, 0.01% poly histidine, and 0.01% poly-L-lysine. At least three separate runs were conducted with each solution, and the solution was thoroughly washed in between the measurements. The measured I-V data was exported to Microsoft Excel for fitting in the analytical model.

## 3. Results

### 3.1 EGFET analytical model incorporating mobility degradation and series resistance:

Here the analytical model for the EGFET I-V characteristics is derived in the linear region of transistor operation. The following nomenclature is used: $C_{ox}$ is the gate oxide capacitance, $\mu_o$ is the low-field mobility, W and L are the width and length respectively, $V_T$ is the threshold voltage, $\theta$ is the surface roughness parameter, $R_s$ is the source resistance, and $R_{sd}$ is the combined source and drain resistances. The extrinsic gate-to-source ($V_{gs}$) and extrinsic drain-to-source voltages ($V_{ds}$) are related to the intrinsic gate-to-source ($V_{GS}$) and intrinsic drain-to-source voltages ($V_{DS}$) as:

$$V_{GS} = V_{gs} - I_{ds}R_s \tag{1}$$

$$V_{DS} = V_{ds} - I_{ds}R_{sd} \tag{2}$$

Because the parasitic resistances, $R_s$ and $R_d$, are inherently tied up with the EGFET terminal voltages, they need to be determined independently before extracting the other parameters from the I-V characteristics. Using the gradual channel approximation and plugging in equations (1-2), the EGFET drain current can be written as [10,11]:

$$I_{ds} = \frac{\mu_o(W/L)C_{OX}\,(V_{gs}-I_{ds}R_s-V_T)\,(V_{ds}-I_{ds}R_{sd})}{1+\theta(V_{gs}-I_{ds}R_s-V_T)} \tag{3}$$

where the effective mobility is given by [10,11]:

$$\mu_{eff} = \frac{\mu_o}{1+\theta(V_{gs}-I_{ds}R_s-V_T)} \tag{4}$$

Equation (3) incorporates the effects of the surface roughness parameter $\theta$ and the series resistances at the source and drain contacts, $R_s$ and $R_d$. We rewrite equation (3) as a quadratic expression to obtain:

$$I_{ds}^2[R_s\theta + \beta_o R_s R_{sd}] - I_{ds}\left[(V_{gs}-V_T)\beta_o R_{sd} + \beta_o R_s V_{ds} + \left(1+\theta(V_{gs}-V_T)\right)R_s\right] + \beta_o V_{ds}(V_{gs}-V_T) = 0 \tag{5}$$

where $\beta_o = \mu_o C_{OX}(W/L)$.

Assuming $(V_{gs}-V_T)\beta_o R_{sd} > \beta_o R_s V_{ds}$ in the linear region of operation, we solve equation (5) for the EGFET drain current as:

$$I_{ds} = \frac{[1+(\theta+\beta_o R_{sd})(V_{gs}-V_T)] \pm \sqrt{[1+(\theta+\beta_o R_{sd})(V_{gs}-V_T)]^2 - 4R_s(\theta+\beta_o R_{sd})\beta_o V_{ds}(V_{gs}-V_T)}}{2R_s(\theta+\beta_o R_{sd})} \tag{6}$$

Simplifying the square-root term in equation (6), we get:

$$I_{ds} = \frac{[1+(\theta+\beta_o R_{sd})(V_{gs}-V_T)]}{2R_s(\theta+\beta_o R_{sd})}\left\{1 - \sqrt{1 - \frac{4R_s(\theta+\beta_o R_{sd})\beta_o V_{ds}(V_{gs}-V_T)}{[1+(\theta+\beta_o R_{sd})(V_{gs}-V_T)]^2}}\right\} \tag{7}$$

Applying Taylor series expansion, equation (7) reduces to:

$$I_{ds} = \frac{\beta_o V_{ds}(V_{gs}-V_T)}{1+(\theta+\beta_o R_{sd})(V_{gs}-V_T)} \tag{8}$$

The gate transconductance $g_m$ is derived from equation (8) as:

$$g_m = \frac{\partial I_{ds}}{\partial V_{gs}} = \frac{\beta_o V_{ds}(V_{gs}-V_T)}{1+(\theta+\beta_o R_{sd})(V_{gs}-V_T)} - \frac{\beta_o V_{ds}(V_{gs}-V_T)(\theta+\beta_o R_{sd})}{[1+(\theta+\beta_o R_{sd})(V_{gs}-V_T)]^2} \tag{9}$$

Rearranging the terms, equation (9) yields:

$$g_m = \frac{\beta_o V_{ds}}{[1+(\theta+\beta_o R_{sd})(V_{gs}-V_T)]^2} \quad (10)$$

Equation (8) and equation (10) are the final expressions for the drain current $I_{ds}$ and gate transconductance $g_m$ from which we can extract the key parameters – threshold voltage $V_T$, low-field mobility $\mu_0$, surface roughness parameter $\theta$, effective mobility $\mu_{eff}$, and series resistance $R_{SD}$.

As observed from equation (10), the ideal limit of gate transconductance $g_m$ is $\beta_o V_{ds}$ in the absence of mobility degradation parameter ($\theta = 0$) and parasitic series resistance ($R_{sd} = 0$). Plotting the gate transconductance $g_m$ versus gate voltage $V_{gs}$ provides information about the transition from weak inversion to strong inversion. While approaching the threshold voltage from the weak inversion regime, there is a sharp rise in both the gate transconductance $g_m$ and the net inversion charge $Q_{inv}$. Fig. 2a shows the current-gate voltage (I-$V_{gs}$) characteristics and gate transconductance $g_m$ (= $\partial I_{ds} / \partial V_{gs}$) for the EGFET at $V_{ds} = 0.4$ V with deionized (DI) water on the sensing area. The peak of the transconductance curve ($g_{m\_max}$) occurs at the control gate voltage $V_{gs} = 2.2$ V from where the I-V curve is extrapolated backwards to give the threshold voltage $V_T$ at the intercept of the x-axis. Fig. 2b shows the current-drain voltage (I-$V_{ds}$) characteristics and the drain transconductance ($g_{ds} = \partial I_{ds} / \partial V_{ds}$) plots at discrete gate voltages ($V_{gs} = 2$V to 4.5V).

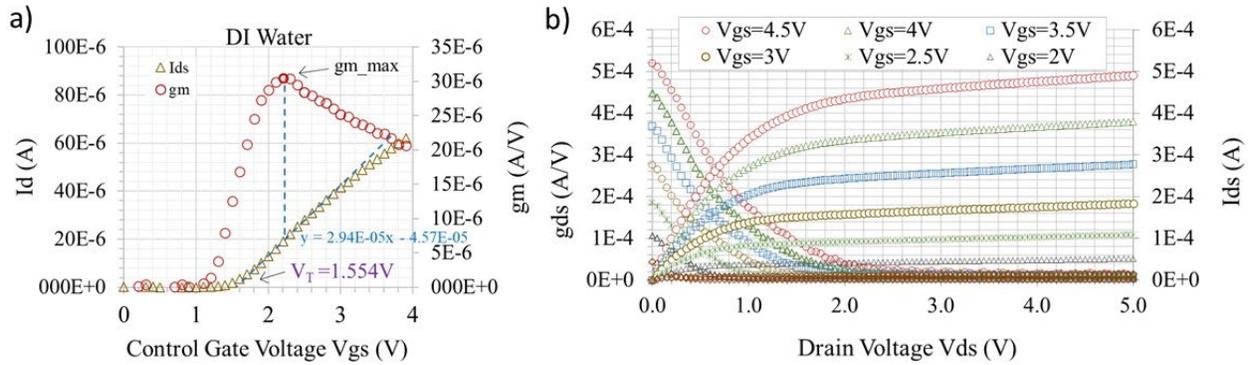

**Fig. 2.** Current-voltage (I-V), gate transconductance, and drain transconductance of EGFET with DI water on the sensing area. (a) The drain current $I_{ds}$ and gate transconductance $g_m$ (where $g_m = \partial I_{ds} / \partial V_{gs}$) are plotted with respect to the gate voltage $V_{gs}$. The drain voltage $V_{ds} = 0.4$ V. (b) The drain current $I_{ds}$ and drain transconductance $g_{ds}$ (where $g_{ds} = \partial I_{ds} / \partial V_{ds}$) are plotted with respect to the drain voltage $V_{ds}$ at different gate voltages $V_{gs}$.

3.2 Peak transconductance method for threshold voltage extraction:

The peak transconductance method is one approach to estimate the threshold voltage $V_T$ of MOSFET devices [6]. In this method, threshold voltage $V_T$ is defined as the gate voltage $V_{gs}$ where the rate of transconductance change ($\partial g_m / \partial V_{gs}$) with respect to the gate voltage $V_{gs}$ exhibits a maximum. At this gate voltage, the surface potential $\phi_s$ is close to the classical $2\phi_F$ point [6,12]. In Fig. 2a using the peak transconductance method, the threshold voltage is estimated to be $V_T = 1.554$ V. Apart from the threshold voltage $V_T$, it is difficult to use the peak transconductance method to directly extract the surface roughness parameter $\theta$ and the series resistance $R_{sd}$. To do this, equations (8, 10) need to be further modified to first extract the threshold voltage $V_T$ and low-field mobility $\mu_o$, and then use the computed values to estimate the surface roughness parameter $\theta$ as described in the sections below.

### 3.3 $I_{ds} / g_m^{0.5}$ method for EGFET parameter extraction:

This method was originally proposed for the extraction of MOSFET device parameters [10] but has not been tested on EGFETs. This method involves calculating the function $I_{ds} / g_m^{0.5}$ using equations (8, 10) as given below:

$$\frac{I_{ds}}{g_m^{0.5}} = \sqrt{\beta_o V_{ds}}\ (V_{gs} - V_T) = \sqrt{\mu_o \frac{W}{L} C_{OX} V_{ds}}\ (V_{gs} - V_T) \tag{11}$$

From equation (11), we notice that $I_{ds} / g_m^{0.5}$ is linearly dependent on the gate voltage $V_{gs}$, while the effects of surface roughness parameter $\theta$ and series resistance $R_{sd}$ are eliminated in this function. Plotting $I_{ds} / g_m^{0.5}$ versus the gate voltage $V_{gs}$ yields a straight line giving the threshold voltage $V_T$ as the x-intercept and the low-field mobility $\mu_o$ from the slope. Using equations (8, 10), the surface roughness parameter $\theta$ is:

$$\theta = \frac{\left[\frac{I_{ds}}{g_m(V_{gs}-V_T)} - 1\right]}{(V_{gs}-V_T)} - \beta_o R_{sd} \tag{12}$$

Knowing the low-field mobility $\mu_o$ and the surface roughness parameter $\theta$, the effective mobility is computed from equation (4). Fig. 3a shows the $I_{ds} / g_m^{0.5}$ method of parameter extraction using the I-V curves from Fig. 2a with DI water on the sensing area. The x-intercept of the fitting line gives the threshold voltage as $V_T = 1.618$ V. The low-field mobility $\mu_o$ is 505 cm²/s using $V_{ds} = 0.4$ V, $W / L = 3$, and $C_{ox} = 57.8 \times 10^{-9}$ cm⁻². The series resistance $R_{sd}$ is measured as the output resistance at the drain terminal $V_{ds}$ which converges to 50 Ω in our case. Fig. 3b plots for surface roughness parameter θ and the effective mobility $\mu_{eff}$. The surface roughness parameter $\theta$ is in the range of 0.11 – 0.13 V⁻¹. The effective mobility is not a constant, but rather degrades with increasing gate voltage $V_{gs}$ as suggested by equation (4).

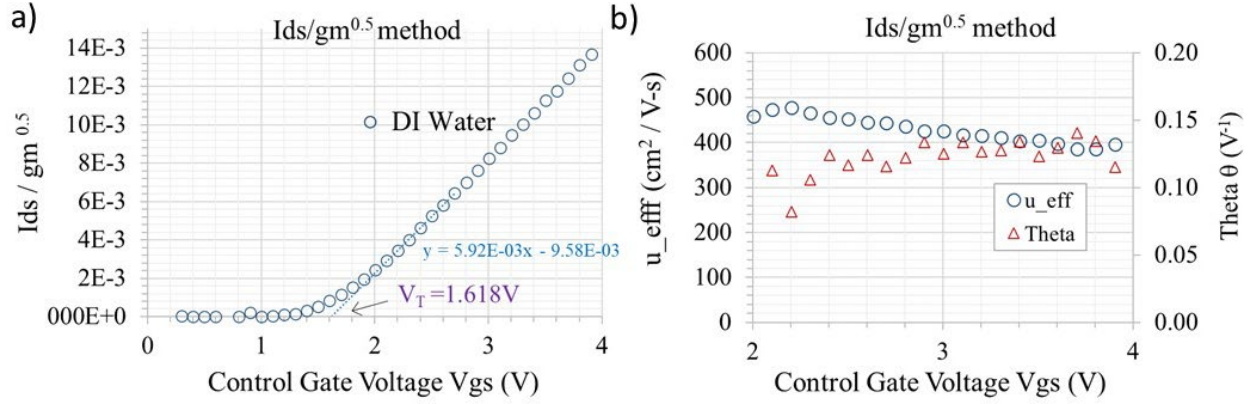

**Fig. 3.** The $I_{ds} / g_m^{0.5}$ method for EGFET parameter extraction with DI water on the sensing area. (a) The function $I_{ds} / g_m^{0.5}$ is plotted with respect to the gate voltage $V_{gs}$ using the I-V data from Fig. 2a. The drain voltage $V_{ds} = 0.4$ V, $W/L = 3$, and $C_{ox} = 57.8 \times 10^{-9}$ cm$^{-2}$. The fitted line gives a threshold voltage $V_T = 1.618$ V and a low-field mobility $\mu_o = 505$ cm$^2$/s. (b) Using the known values from Fig. 3a, the surface roughness parameter θ and the effective mobility $\mu_{eff}$ are plotted as a function of the gate voltage $V_{gs}$.

### 3.4 1/$I_{ds}$ method for EGFET parameter extraction:

This proposed method involves computing the second derivative of the drain current $I_{ds}$ with respect to the gate voltage $V_{gs}$ (i.e., $\partial^2(1/I_{ds}) / \partial V_{gs}^2$) to eliminate the effects of surface roughness parameter θ and the series resistance $R_{sd}$. In this method, equation (8) is re-written as:

$$\frac{1}{I_{ds}} = \frac{1}{\beta_o V_{ds}(V_{gs}-V_T)} + \frac{(\theta + \beta_o R_{sd})}{\beta_o V_{ds}} \quad (13)$$

Equation (13) is differentiated twice with respect to $V_{gs}$ to give:

$$\frac{\partial(1/I_{ds})}{\partial V_{gs}} = -\frac{1}{\beta_o V_{ds}(V_{gs}-V_T)^2} \quad (14)$$

$$\frac{\partial^2(1/I_{ds})}{\partial V_{gs}^2} = \frac{2}{\beta_o V_{ds}(V_{gs}-V_T)^3} \quad (15)$$

Equation (15) is re-arranged as a straight-line function:

$$\left[\frac{\partial^2(1/I_{ds})}{\partial V_{gs}^2}\right]^{-1/3} = \left[\frac{2}{\beta_o V_{ds}}\right]^{-1/3} (V_{gs} - V_T) \quad (16)$$

From equation (16), we see that the plot of the inverse cube root of the function $\partial^2(1/I_{ds}) / \partial V_{gs}^2$ with respect to the gate voltage $V_{gs}$ gives the threshold voltage $V_T$ as the x-intercept and the low-field mobility $\mu_o$ from the slope of the straight line. Fig. 4a demonstrates the 1/$I_{ds}$ method of parameter extraction using the I-V curves from Fig. 2a with DI water on the sensing area. The x-intercept of the fitting line gives the threshold voltage as $V_T = 1.557$ V. The low-field mobility $\mu_o$ is 574 cm²/s using $V_{ds} = 0.4$ V, $W/L = 3$, and $C_{ox} = 57.8 \times 10^{-9}$ cm⁻². Fig. 4b plots for surface roughness parameter $\theta$ and the effective mobility $\mu_{eff}$. The surface roughness parameter $\theta$ is in the range of $0.05 - 0.3$ V⁻¹. The range of surface roughness parameter $\theta$ here (Fig. 4b) is relatively larger than that from the $I_{ds}/g_m^{0.5}$ method (Fig. 3b) because of the noise generated from the calculation of the second derivative of the inverse drain current. The generated noise involved in calculating the second derivative of the inverse current is also apparent in Fig. 4a which tends to increase in the strong inversion region.

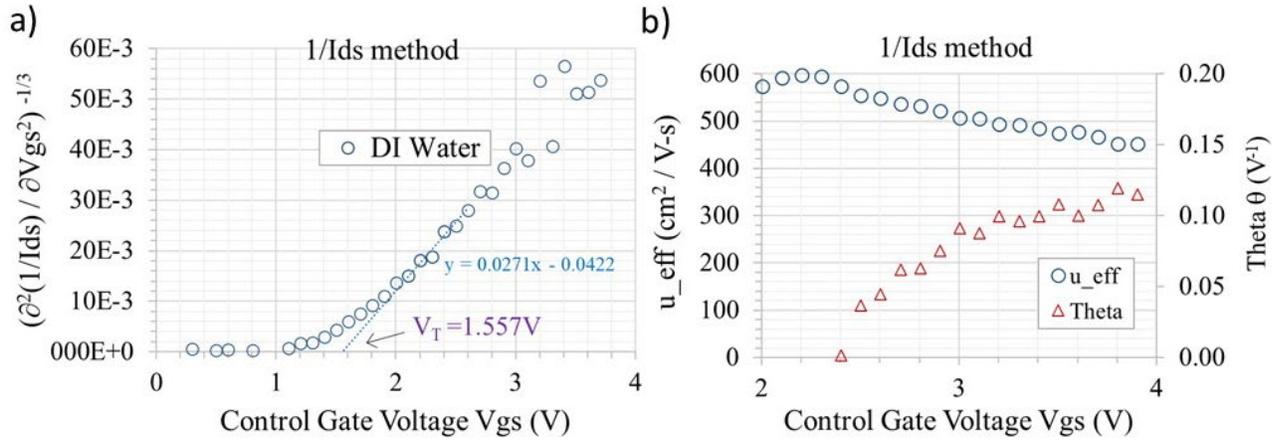

**Fig. 4.** The 1/$I_{ds}$ method for EGFET parameter extraction with DI water on the sensing area. (a) The function $[\partial^2(1/I_{ds})/\partial V_{gs}^2]^{-1/3}$ is plotted with respect to the gate voltage $V_{gs}$ using the I-V data from Fig. 2a. The drain voltage $V_{ds} = 0.4$ V, $W/L = 3$, and $C_{ox} = 57.8 \times 10^{-9}$ cm⁻². The fitted line gives a threshold voltage $V_T = 1.557$ V and a low-field mobility $\mu_o = 574$ cm²/s. (b) Following Fig. 4a, the surface roughness parameter θ and the effective mobility μ$_{eff}$ are plotted as a function of the gate voltage $V_{gs}$.

3.5 $g_{ds}/(\partial g_{ds}/\partial V_{gs})^{0.5}$ method for EGFET parameter extraction:

This proposed method involves calculating the drain conductance $g_{ds}$ ($= \partial I_{ds}/\partial V_{ds}$) and its derivative with respect to the gate voltage $V_{gs}$, and eventually calculating the function $g_{ds}/(\partial g_{ds}/\partial V_{gs})^{0.5}$. From equation (8), the drain conductance $g_{ds}$ and its reciprocal are obtained as:

$$g_{ds} = \frac{\partial I_{ds}}{\partial V_{ds}} = \frac{\beta_o(V_{gs}-V_T)}{1+(\theta+\beta_o R_{sd})(V_{gs}-V_T)} \tag{17}$$

$$\frac{1}{g_{ds}} = \frac{1}{\beta_o(V_{gs}-V_T)} + \frac{(\theta+\beta_o R_{sd})}{\beta_o} \tag{18}$$

From equations (17, 18), the derivative of $g_{ds}$ with respect to the gate voltage $V_{gs}$ is expressed as:

$$\frac{\partial g_{ds}}{\partial V_{gs}} = \frac{g_{ds}^2}{\beta_o(V_{gs}-V_T)^2} \tag{19}$$

From equations (17, 19), the function $g_{ds} / (\partial g_{ds}/ \partial V_{gs})^{0.5}$ is written as:

$$\frac{g_{ds}}{\left(\frac{\partial g_{ds}}{\partial V_{gs}}\right)^{0.5}} = \sqrt{\beta_o}(V_{gs} - V_T) = \sqrt{\mu_o \frac{W}{L} C_{OX}}(V_{gs} - V_T) \tag{20}$$

Equation (20) indicates that the plot of the function $g_{ds} / (\partial g_{ds}/ \partial V_{gs})^{0.5}$ with respect to the gate voltage $V_{gs}$ is a straight line where its x-intercept gives the threshold voltage $V_T$ and its slope gives the low-field mobility $\mu_o$. Fig. 5a demonstrates the $g_{ds} / (\partial g_{ds}/ \partial V_{gs})^{0.5}$ method of parameter extraction using the I-V curves from Fig. 2a with DI water on the sensing area. Using $W / L = 3$, and $C_{ox} = 57.8 \times 10^{-9}$ cm$^{-2}$ at $V_{ds} = 0.4$ V, the threshold voltage is $V_T = 1.257$ V and the low-field mobility $\mu_o$ is 316 cm$^2$/s. The threshold voltage decreases with increasing drain voltage, as shown earlier in MOSFET devices [11]. Plugging these values in equation (8), the surface roughness parameter $\theta$ is obtained in the range of 0.01 – 0.2 V$^{-1}$ (Fig. 5b).

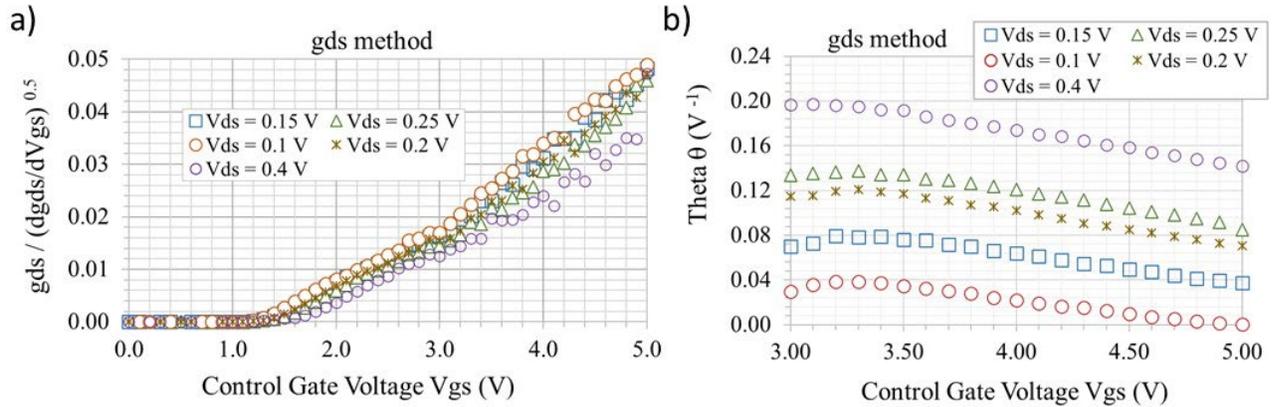

**Fig. 5.** The $g_{ds} / (\partial g_{ds}/ \partial V_{gs})^{0.5}$ method for EGFET parameter extraction. (a) The function $g_{ds} / (\partial g_{ds}/ \partial V_{gs})^{0.5}$ is plotted versus gate voltage $V_{gs}$ at different drain voltages $V_{ds}$. (b) The surface roughness parameter $\theta$ is extracted after calculating the threshold voltage $V_T$ and low-field mobility $\mu_o$ from equation (20) using $W / L = 3$, and $C_{ox} = 57.8 \times 10^{-9}$ cm$^{-2}$.

### 3.6 Testing with charged polyelectrolytes:

Here our goal is to apply the EGFET parameter extraction methods on I-V experiments having positive or negative charges on the sensing area, and to compare the resulting shift in threshold voltage from the different methods (Fig. 3-5). Three different polyelectrolytes are tested on the sensing area of EGFET – 0.5 % poly aspartic acid, 0.01% poly-L-lysine, and 0.01% poly histidine. Fig. 6a shows the skeletal formula of poly aspartic acid (pKa = 3.90) having carboxylic groups that contribute to its negative charge [13,14]. Fig. 6b plots the I-V characteristics and gate transconductance $g_m$ at $V_{ds}$ = 0.4 V. Using the peak transconductance method, the maximum transconductance ($g_{m\_max}$) occurs at the gate voltage $V_{gs}$ = 2.2 V which gives the threshold voltage extrapolated from the x-intercept as $V_T$ = 1.696 V. The $I_{ds} / g_m^{0.5}$ method gives the threshold voltage $V_T$ = 1.696 V (Fig. 6c). The low-field mobility $\mu_o$ is 391 cm$^2$/s while the surface roughness parameter $\theta$ is in the range of 0.08 – 0.125 V$^{-1}$. The 1/$I_{ds}$ method gives the threshold voltage as $V_T$ = 1.705 V (Fig. 6d). The low-field mobility $\mu_o$ is 647 cm$^2$/s while the surface roughness parameter $\theta$ is in the range of 0.1 – 0.13 V$^{-1}$.

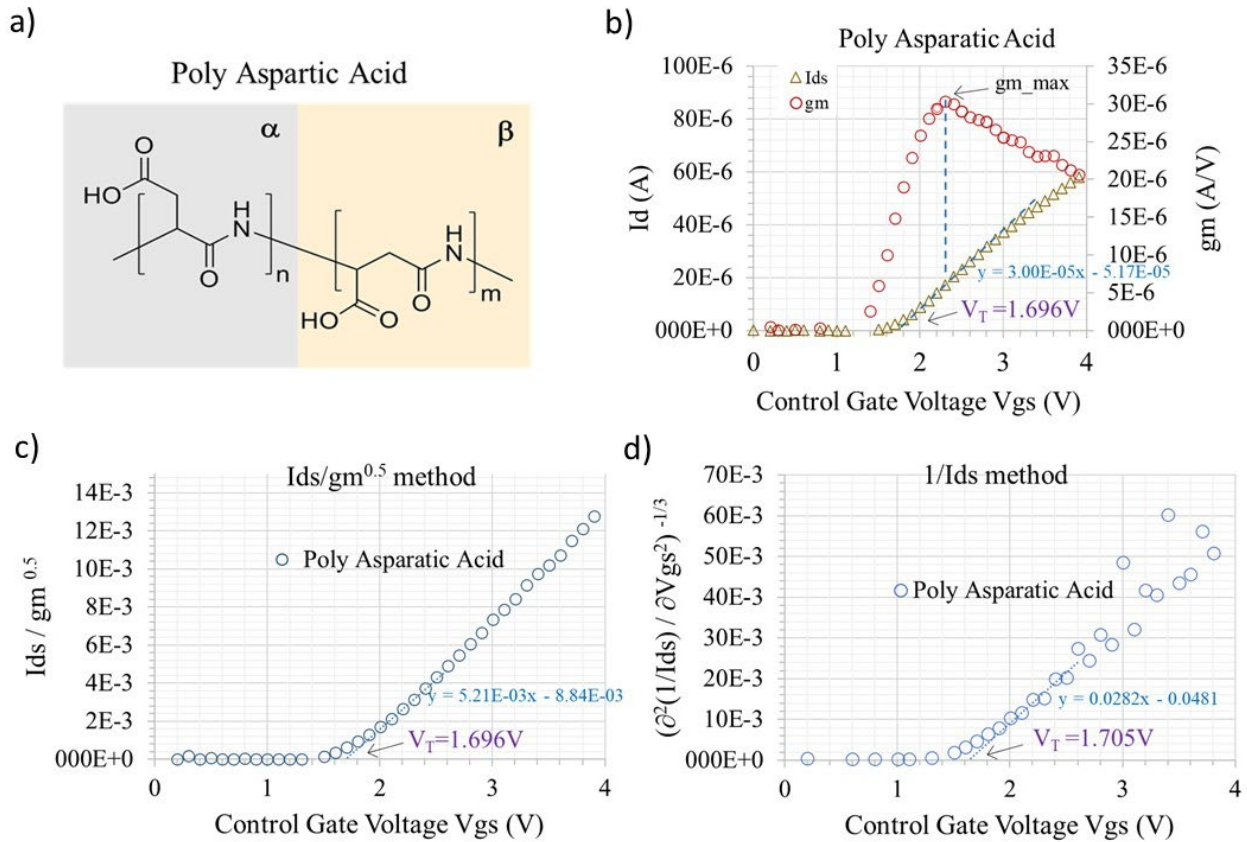

**Fig. 6.** EGFET parameter extraction with poly aspartic acid (negatively charged) on the sensing area. (a) Poly aspartic acid has a protein-like amide bond as its backbone that offers biodegradability and a carboxylic acid group that offers its anionic property [13,14]. (b) The drain current $I_{ds}$ and gate transconductance $g_m$ are plotted with respect to the gate voltage $V_{gs}$. The extracted threshold voltage is $V_T$ = 1.696 V (c) The function $I_{ds} / g_m^{0.5}$ is plotted with respect to the gate voltage $V_{gs}$ using the I-V data from Fig. 6b. The fitted line gives a threshold voltage $V_T$ = 1.696 V and a low-field mobility $\mu_o$ = 391 cm$^2$/s. The surface roughness parameter $\theta$ is in the range of 0.08 – 0.125 V$^{-1}$. (d) The function $[\partial^2(1/I_{ds}) / \partial V_{gs}^2]^{-1/3}$ is plotted with respect to the gate voltage $V_{gs}$ using the I-V data from Fig. 6b. The fitted line gives a threshold voltage $V_T$ = 1.705 V and a low-field mobility $\mu_o$ = 647 cm$^2$/s. The surface roughness parameter $\theta$ is in the range of 0.1 – 0.13 V$^{-1}$. In the above calculations, the drain voltage $V_{ds}$ = 0.4 V, $W/L$ = 3, and $C_{ox}$ = 57.8 × 10$^{-9}$ cm$^{-2}$.

Fig. 7a shows the skeletal formula of poly-L-Lysine (pKa = 9.85) having a hydrophilic amino group that makes it weakly positively charged and is often used for coating solid surfaces to promote adhesion to negatively charged cell membranes [15]. Fig. 7b plots the I-V characteristics and gate transconductance $g_m$ at $V_{ds}$ = 0.4 V. Using the peak transconductance method, the maximum transconductance ($g_{m\_max}$) occurs at the gate voltage $V_{gs}$ = 2.2 V giving a threshold voltage of $V_T$ = 1.505 V. Fig. 7c shows the $I_{ds} / g_m^{0.5}$ method which gives the threshold voltage as $V_T$ = 1.541 V. The low-field mobility $\mu_o$ is 495 cm²/s while the surface roughness parameter $\theta$ is in the range of 0.09 – 0.105 V$^{-1}$. Fig. 7d shows the $1/I_{ds}$ method which gives the threshold voltage as $V_T$ = 1.50 V. The low-field mobility $\mu_o$ is 580 cm²/s while the surface roughness parameter $\theta$ is in the range of 0.09 – 0.12 V$^{-1}$.

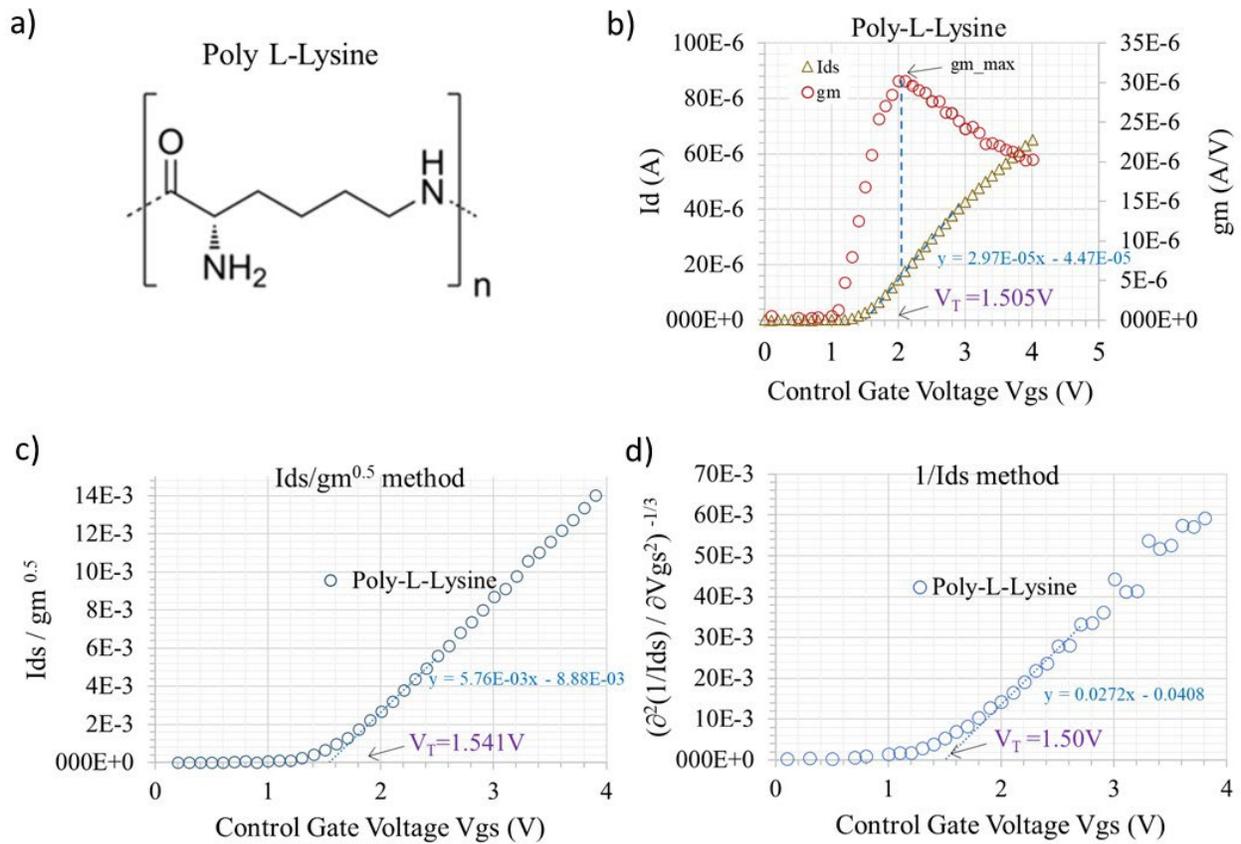

**Fig. 7.** EGFET parameter extraction with poly-L-Lysine (positively charged) on the sensing area. (a) Poly-L-Lysine has hydrophilic amino groups and provides positively charge on the sensing area [16]. It is often used to promote surface adhesion to negatively charged cell membranes. (b) The drain current $I_{ds}$ and gate transconductance $g_m$ are plotted with respect to the gate voltage $V_{gs}$. The extracted threshold voltage is $V_T$ = 1.505 V (c) The function $I_{ds} / g_m^{0.5}$ is plotted with respect to the gate voltage $V_{gs}$ using the I-V data from Fig. 7b. The fitted line gives a threshold voltage $V_T$ = 1.541 V and a low-field mobility $\mu_o$ = 494 cm²/s. The surface roughness parameter $\theta$ is in the range of 0.09 – 0.105 V$^{-1}$. (d) The function $[\partial^2(1/I_{ds}) / \partial V_{gs}^2]^{-1/3}$ is plotted with respect to the gate voltage $V_{gs}$ using the I-V data from Fig. 6b. The fitted line gives a threshold voltage $V_T$ = 1.50 V and a low-field mobility $\mu_o$ = 580 cm²/s. The surface

roughness parameter θ is in the range of 0.09 – 0.12 V$^{-1}$. In the above calculations, the drain voltage $V_{ds}$ = 0.4 V, $W/L$ = 3, and $C_{ox}$ = 57.8 × 10$^{-9}$ cm$^{-2}$.

Fig. 8a shows the skeletal formula of poly Histidine (pKa = 6.04) that has a protonated NH$_3^+$ conferring its positive charge [17]. Fig. 8b plots the I-V characteristics and gate transconductance $g_m$ at $V_{ds}$ = 0.4 V. Using the peak transconductance method, the maximum transconductance ($g_{m\_max}$) occurs at the gate voltage $V_{gs}$ = 2.2 V giving a threshold voltage as $V_T$ = 1.576 V. Fig. 8c shows the $I_{ds}/g_m^{0.5}$ method which gives the threshold voltage as $V_T$ = 1.627 V. The low-field mobility $\mu_o$ is 709 cm$^2$/s and the surface roughness parameter θ is in the range of 0.2 – 0.35 V$^{-1}$. Fig. 8d shows the 1/$I_{ds}$ method of parameter extraction gives the threshold voltage as $V_T$ = 1.543 V. The low-field mobility $\mu_o$ is 634 cm$^2$/s and the surface roughness parameter θ is in the range of 0.15 – 0.35 V$^{-1}$.

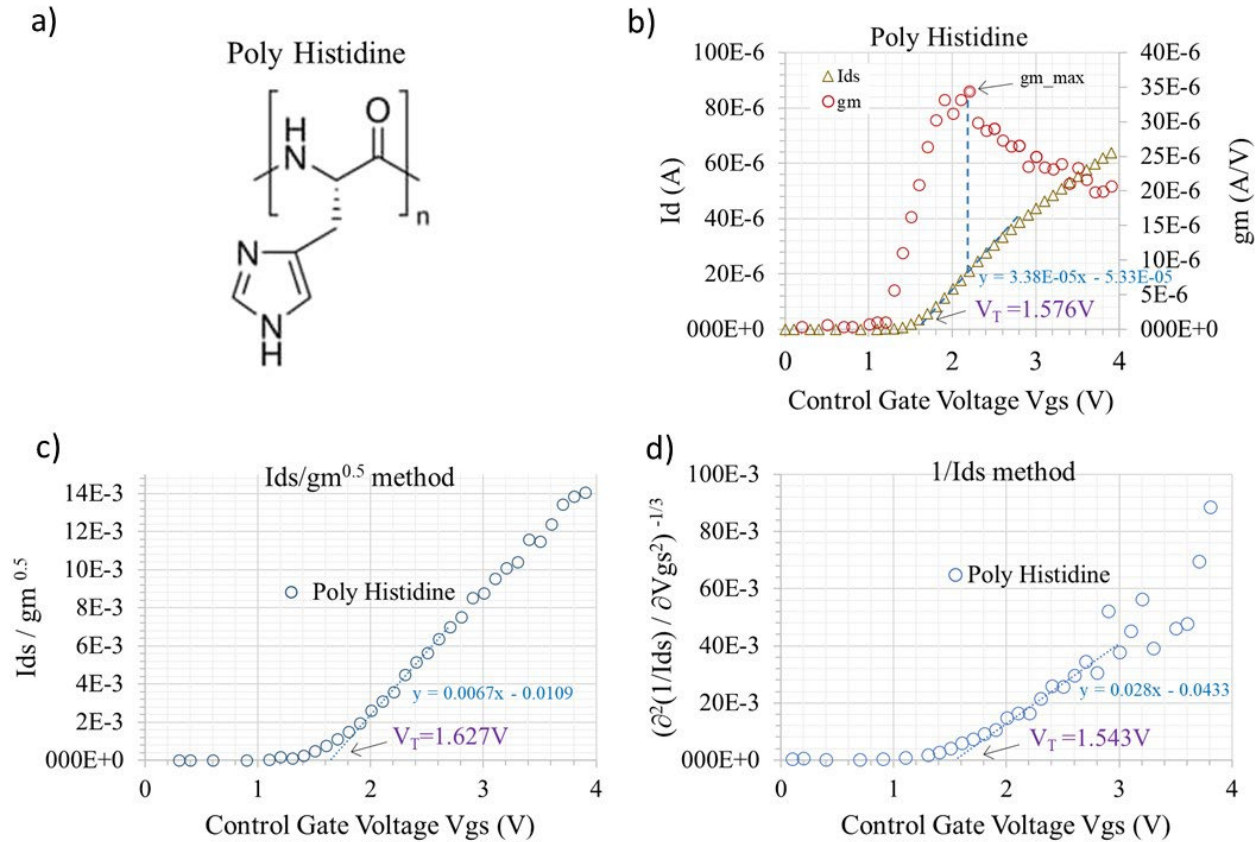

**Fig. 8.** EGFET parameter extraction with poly Histidine (positively charged) on the sensing area. (a) Poly Histidine contains a protonated amino group, a carboxylic acid group and an imidazole side chain [18]. (b) The drain current $I_{ds}$ and gate transconductance $g_m$ (where $g_m = \partial I_{ds}/\partial V_{gs}$) are plotted with respect to the gate voltage $V_{gs}$. The extracted threshold voltage is $V_T$ = 1.576 V (c) The function $I_{ds}/g_m^{0.5}$ is plotted with respect to the gate voltage $V_{gs}$ using the I-V data from Fig. 8b. The fitted line gives a threshold voltage $V_T$ = 1.627 V and a low-field mobility $\mu_o$ = 709 cm$^2$/s. The surface roughness parameter θ is in the range of 0.2 – 0.35 V$^{-1}$. (d) The function $[\partial^2(1/I_{ds}) / \partial V_{gs}^2]^{-1/3}$ is plotted with respect to the gate voltage $V_{gs}$ using the I-V data from Fig. 6b. The fitted line gives a threshold voltage $V_T = 1.543$ V and a low-field mobility $\mu_o = 634$ cm$^2$/s. The surface roughness parameter $\theta$ is in the range of 0.15 – 0.35 V$^{-1}$. In the above calculations, the drain voltage $V_{ds} = 0.4$ V, $W / L = 3$, and $C_{ox} = 57.8 \times 10^{-9}$ cm$^{-2}$.

## 4. Discussion

The choice of an appropriate model for EGFET parameter extraction is dependent on the tradeoffs between the simplicity of chosen method and the accuracy of extracted parameter. To reiterate this point, the linear extrapolation technique or the constant current technique are simple threshold voltage extraction methods but do not incorporate mobility degradation effects [1,3,4]. On the other end, physics-based numerical simulations are more accurate but present a steep learning curve to grasp its full potential [19,20]. Here we presented EGFET parameter extraction methods based on closed-form analytical models, standard I-V measurements, and straight-line extrapolation of the involved functions. One benefit of the presented analytical models is the intuitive understanding of different parasitic effects on the EGFET I-V characteristics, which further helps to improve the device performance beyond the Nernst potential. From our results, we observed that the generated noise is significant (in the strong inversion region) for the method based on the second derivative function (i.e., $\partial^2(1/I_{ds}) / \partial V_{gs}^2$), which results in a broader range of surface roughness parameter $\theta$ values. For the methods based on gate transconductance $g_m$, the extracted threshold voltage $V_T$ was relatively consistent while the low-field mobility $\mu_o$ was also in an acceptable range 505-709 cm$^2$/s. The method based on the drain conductance $g_{ds}$ and the function $g_{ds} / (\partial g_{ds} / \partial V_{gs})^{0.5}$ yielded lower values of the threshold voltage $V_T$ and low-field mobility $\mu_o$ compared to those based on the gate transconductance $g_m$ methods. To further validate the threshold voltage $V_T$ extraction methods, capacitance-voltage (C-V) measurements could be performed on the EGFETs. This is because the inversion capacitance sharply rises near the threshold voltage causing the gate-to-channel capacitance to approach the net oxide capacitance, and the gate-to-channel capacitance converges to the oxide capacitance at the threshold voltage [21,22].

While comparing the threshold voltage shifts, the effective mobility of EGFETs is also an important parameter to consider for new device structures, such as those fabricated on organic substrates [9,23]. The effective mobility is not constant in the strong inversion region but is rather influenced by the surface roughness parameter $\theta$ and series resistance $R_{sd}$. This mobility degradation effect is often ignored in the EGFET parameter extraction as discussed in Section 1. The surface roughness parameter $\theta$ is a measure of the smoothness of the electron transport surface, such as silicon/silicon oxide (Si/SiO$_2$) interface. A rough Si/SiO$_2$ interface produces a higher $\theta$ value, a lower critical electric field, and subsequently a lowered peak transconductance value. Increasing gate voltage also tends to increase the $\theta$ value, which contributes to the mobility degradation in strong inversion. In addition, the series resistance $R_{sd}$ produces an effective increase in the surface roughness parameter $\theta$ by a factor $\beta_o R_{ds}$. In other words, the effective surface roughness parameter is increased from $\theta$ to $\theta + \beta_o R_{ds}$ due to the series resistance $R_{sd}$ as indicated by equation (12). It follows that a higher value of series resistance $R_{sd}$ causes a more significant mobility degradation of EGFETs. As such, it is beneficial to extract the values of $\theta$ and $R_{sd}$ from different methods to compare their relative values in the voltage range of operation.

For the extraction of the series resistances, $R_s$ and $R_d$, there are different I-V methods proposed in the MOSFET community [24,25] which could be tested for EGFETs, especially in devices having external sensors, long leads, or new source/drain junction contacts [26–32]. In one approach adopted for MOSFETs, the total channel resistance $R_{ch}$ is expressed as a ratio of drain voltage $V_{ds}$ to the drain current $I_{ds}$ [33,34]. Following equation (8), we can write the total channel resistance as:

$$R_{ch} = \frac{V_{ds}}{I_{ds}} = \frac{L_{mask} - \Delta L}{\mu_o W C_{OX}(V_{gs} - V_T)} + \frac{(\theta + \beta_o R_{sd})}{\beta_o} \tag{21}$$

where $\Delta L$ is the difference between the mask channel length $L_{mask}$ and the effective channel length. To extract the total series resistance $R_{sd}$ from equation (21), the I-V characteristics are measured for several devices (having different mask channel lengths) operating in the linear region at different gate voltages $V_{gs}$. The channel resistance $R_{ch}$ is plotted as a function of varying mask channel lengths $L_{mask}$. Assuming the mobility degradation is negligible, the intersection of all the straight lines gives the value of the total parasitic series resistance $R_{sd}$. This method requires the availability of multiple devices of different mask channel lengths, which may not be practical for users having a single device. In our case, we chose to directly measure the output resistance of a single device. While our method is relatively less demanding in measurement, it only gives the total series parasitic resistance $R_{sd}$ and does not provide separate values of $R_s$ and $R_d$. As such, our method is inappropriate in the case of asymmetric devices where $R_s \neq R_d$ or in situations where $R_s$ and $R_d$ are bias dependent. Numerical simulations have shown that differences in the values of $R_s$ and $R_d$ primarily arise from the asymmetry in the contact resistances at the source and drain terminals, and not from the differences in doping densities at the source and drain regions or gate misalignment [35,36].

**Conclusion**

The importance of accurate EGFET device characterization goes beyond measuring the threshold voltage of transistors. It is also valuable to extract the effective mobility and series resistance of the source/drain contacts to accurately evaluate and improve the EGFET device performance. In the strong inversion region, these parasitic effects become more significant, and as such, it is advisable to extract the key parameters in the linear region of operation where the low-field mobility is relatively constant. With this rationale, we derived an analytical model for EGFETs in the linear region of operation while incorporating the effects of surface roughness and series resistance. Different functions (involving the drain current, gate transconductance, and drain transconductance) were derived which resulted in straight-line plots yielding the key parameters from the slope and intercept. The gate transconductance methods produced nearly consistent values of the threshold voltage, low-field mobility, and effective mobility, whereas the drain transconductance method gave relatively lower values of these parameters. In the experiments with polyelectrolytes, it is confirmed that negative charges on the sensing area increase the threshold voltage (i.e., produce a positive $V_T$ shift) while positive charges on the sensing area lower the threshold voltage (i.e., produce a negative $V_T$ shift) from the DI water experiments as reference. Besides the methods discussed here, there are other MOSFET parameter extraction methods which could be tested on EGFETs for a complete comparison of their appropriateness for such charge sensing devices. The effects of body bias and asymmetric source/drain resistance could be further incorporated. It is worth noting that the applicability of any method would depend on its simplicity, accuracy, fabrication costs, measurement ease, device topology, and computational complexity. As such, it is recommended to experiment with all the suggested methods on the fabricated EGFETs to evaluate their relative benefits and pitfalls.

## Acknowledgements

This work was partially supported by U.S. Department of Agriculture to S. P. (grant number: 2023-68008-39857 and 2020-67021-31964).

**Biographies**

Yunsoo Park is a doctoral candidate in the Department of Electrical and Computer Engineering at Iowa State University. He received his Bachelor's degree from South Korea, and served in the South Korea Military for ten years in their engineering division as the Technical Lead. His current research interests are sensor fabrication and characterization, measurement techniques, data analysis, and modeling.

Santosh Pandey is an Associate Professor in the Department of Electrical and Computer Engineering at Iowa State University. He has published several refereed papers with his students and collaborators in the areas of bioengineering, microelectronics, microfluidics, sensors, machine intelligence, plant pathology, electrophysiology, data analytics, and drug screening. After graduating with Bachelor of Technology in Electronic and Electrical Communication Engineering from the Indian Institute of Technology (Kharagpur), he completed his M.S. and Ph.D. in Electrical Engineering from Lehigh University in Bethlehem, Pennsylvania in 2006.

**CRediT authorship contribution statement**

Yunsoo Park: conceptualization, methodology, software, data curation, validation, writing-original draft. Santosh Pandey: project administration, conceptualization, methodology, writing-original draft, resources, software, supervision, reviewing, editing.

**Declaration of Competing Interest**

No potential conflict of interest reported by the authors.